# Quantum-Hall plateau-plateau transition in top-gated epitaxial graphene grown on SiC (0001)

T. Shen,[1,2] A.T. Neal,[1] M.L. Bolen,[1] J.J. Gu,[1] L.W. Engel,[3] M.A. Capano,[1] and P.D. Ye[1,*]

[1]*Electrical and Computer Engineering, Birck Nanotechnology Center, Purdue University, West Lafayette, IN 47907*

[2]*Department of Physics, Birck Nanotechnology Center, Purdue University, West Lafayette, IN 47907*

[3]*National High Magnetic Field Laboratory, Tallahassee, FL 32310*

(Dated: November 15, 2011)

Abstract

We investigate the low-temperature magneto-transport properties of monolayer epitaxial graphene films formed on the Si-face of semi-insulating 4H-SiC substrates by a high temperature sublimation process. A high-$k$ top-gate on the epitaxial graphene is realized by inserting a fully oxidized nanometer thin aluminum film as a seeding layer, followed by an atomic layer deposition process. At low temperatures, the devices demonstrate a strong field effect by the top gate with an on/off ratio of ~7 and an electron mobility up to ~3250 cm$^2$/Vs. After the observation of the half-integer quantum Hall effect for monolayer epitaxial graphene films, detailed magneto-transport measurements have been carried out including varying densities, temperatures, magnetic fields and currents. We study the width of the distinguishable quantum-Hall plateau to plateau transition (Landau level index $n$=0 to $n$=1) as temperature (T) and current are varied. For both gate voltage and magnetic field sweeps and $T$>10 K the transition width goes as $T^{-\kappa}$ with exponent $\kappa$ ~0.42. This universal scaling exponent agrees well with those found in III-V heterojunctions with short range alloy disorders and in exfoliated graphene.

[*]Author to whom correspondence should be addressed. Electronic mail: yep@purdue.edu

## I. INTRODUCTION

Graphene, a single sheet of carbon atoms tightly packed into a two-dimensional (2D) hexagonal lattice, is one of the most exciting electronic materials of recent years. It has been considered a promising candidate for the next generation of high speed electronic devices due to its extraordinary electrical properties, such as a carrier mobility of ~25,000 $cm^2$/Vs and a carrier velocity of ~$10^8$ cm/s at room temperature [1]. Mechanical exfoliation from highly-ordered pyrolytic graphite (HOPG) onto $SiO_2$ could produce small (tens of microns) areas of high quality graphene films [2][3]; however, this exfoliation process cannot form the basis for a large-scale manufacturing process. Recent reports of large-area epitaxial graphene by thermal decomposition of SiC wafers or chemical vapor deposition (CVD) of graphene on Ni or Cu have provided the missing pathway to a viable electronics technology [4-14].

With the advent of these growth techniques driven by electronics industry interests, the opportunity is now available to study the physics of Dirac fermions through engineering graphene-based device structures over large areas, in much the same way as the conventional 2D electron system (2DES) using III-V hetero-junctions or quantum wells. The effects from the substrate and top dielectrics on epitaxial graphene make it a system quite distinct from exfoliated graphene. Therefore, it is important to study the magneto-transport properties of epitaxial graphene films to benchmark with the well-studied exfoliated graphene. The half-integer quantum Hall effect (QHE), a hallmark of the 2D Dirac fermions in graphene, is well-developed in these gated epitaxial graphene samples. Furthermore, their robust plateau-plateau transitions allow study of the possible quantum phase transitions in the epitaxial graphene system.

In this paper, we report detailed studies of the quantum Hall effect of epitaxial graphene films formed on the Si-face of SiC substrate with high-quality top-gate dielectrics. The newly discovered half-integer quantum Hall effect of epitaxial graphene films grown on both the Si-face [10-12] and the C-face [13] of SiC confirm that epitaxial graphene and exfoliated graphene both have Dirac charge carriers. Unlike exfoliated graphene and transferred CVD graphene on a $SiO_2$/Si substrate, which can serve as a global back gate to tune the carrier density of the graphene films, epitaxial graphene can only be charge-density-modulated with a top gate. This property raises the challenge of forming high-quality, ultrathin gate dielectrics with low interface trap density on top of graphene. A perfect graphene surface is chemically inert, which does not lend itself to the conventional atomic layer deposited (ALD) high-$k$ dielectric process [14][15][16]. Here, ALD high-$k$ gate stack integration on epitaxial graphene films is achieved without significant mobility degradation by inserting a fully oxidized ultrathin aluminum film as a seeding layer before ALD process [10,17].

## II. EXPERIMENTAL METHODS

The graphene films were grown on semi-insulating 4H-SiC substrates in an Epigress VP508 SiC hot-wall chemical vapor deposition (CVD) reactor. The off-cut angle of the substrate is nominally zero degrees. Prior to growth, substrates were subjected to a hydrogen etch at 1600°C for 5 minutes, followed by cooling the samples to below 700°C. After evacuating hydrogen from the system, the growth environment is pumped to an approximate pressure of $2 \times 10^{-7}$ mbar before temperature ramping at a rate of 10-20°C/min and up to a specified growth temperature.

The growth conditions, film morphology, and electrical properties of the epitaxial graphene films differ markedly between films grown on the C-face and films grown on the Si-face, as reported by several different groups [18,19,20]. On Si-face, continuous single layer graphene can be formed with a controlled process having a typical Hall mobility of ~1500-2000 $cm^2$/Vs at room temperature. Here, we focus on two types of samples; Samples 1118A, 1118B, 1189A3, and 1189A5 were grown at 1600 °C for 10 minutes in vacuum and sample 1117A grown at 1600°C for 10 minutes in a 10 mbar argon ambient. An additional hydrogen passivation step was applied to samples 1189A3 and 1189A5 following growth. So far, no strong correlation between the film mobility and the growth ambient on the Si-face has been observed, while there is a positive correlation for the C-face found by Tedesco et al [21] and Bolen et al [22].

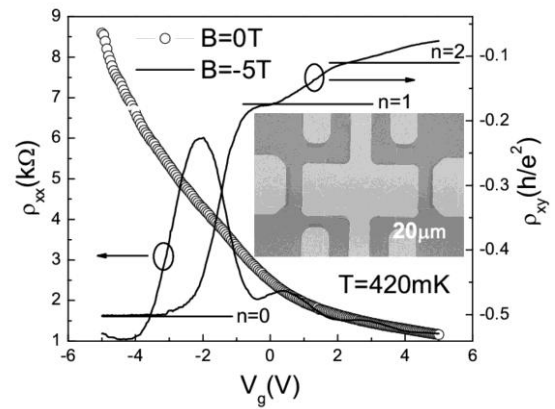

Fig. 1 (a) Four-terminal resistivity $\rho_{xx}$ as a function of top gate bias ($V_g$) of device 1118B measured at 420 mK with zero magnetic field (open circles). Both QHE and SdH oscillations are observed in $\rho_{xy}$ and $\rho_{xx}$ respectively at $B$=-5T (solid lines). Inset: SEM image of a fabricated device. The scale bar is 20 μm.

The detailed device structure is illustrated in the inset of Figure 1. The device isolation was achieved by $O_2$ plasma mesa dry etching. One nanometer of aluminum metal film was evaporated onto the sample by electron-beam evaporation at ~$10^{-6}$ torr and fully oxidized in an oxygen rich ambient for 1 hour or in ambient air overnight as a seeding layer for ALD growth. As the gate dielectric, 30 nm of $Al_2O_3$ was deposited at 300°C using an ASM F-120 reactor with tri-methyl aluminum and water vapor as the precursors. The metal contacts and gate electrodes were subsequently patterned and deposited, both using electron-beam evaporated Ti/Au. For devices 1118A, 1118B, and 1117A, the active device area for magneto-transport has a width of 10 μm and a length of 22 μm, and the gate length is 30 μm. Devices 1189A3 and 1189A5 have an active area width of 10 μm, active area length of 44 μm, and gate length of 56 μm. Four-point magneto-transport measurements are performed in a variable temperature $^3$He cryostat (0.4K to 70K) or $^4$He cryostat (1.5K to 300K) in magnetic fields up to 18 T using standard low frequency lock-in techniques. Additionally, DC measurements are used for the current scaling studies. The external magnetic field ($B$) is applied normal to the graphene plane.

III. RESULTS AND DISCUSSION

A. Half-integer quantum Hall-effect

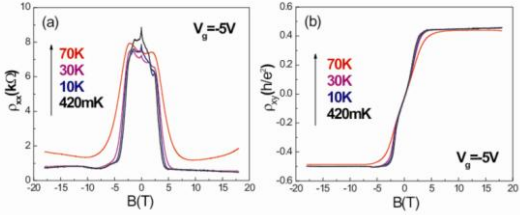

Fig. 2 Temperature dependence of $\rho_{xx}$ and $\rho_{xy}$ at $V_g$= -5V. (a) A pronounced SdH minimum remains up to 70K. The non-symmetric $\rho_{xx}$ along B=0 is ascribed to the imperfection of the device fabrication or inhomogeneous distribution of the charge density in the graphene film. (b) Pronounced n=0 quantum Hall plateau remains up to 70K.

Figure 1 shows four-terminal longitudinal resistivity $\rho_{xx}$ and Hall resistivity $\rho_{xy}$ measured for device 1118B at 420 mK where the top gate voltage is varied between -5V and 5V. At zero magnetic field, $\rho_{xx}$ drops from ~8.6kΩ to ~1.2kΩ, with an on-off ratio of ~7, indicating a good gate-control and the high quality of the top gate dielectric. The increase of $\rho_{xx}$ with the decrease of $V_g$ confirms the initial n-type doping of the graphene channel on SiC (0001). So far, within this bias range, the Fermi level cannot sweep through the charge neutrality point due to the heavy n-type doping induced by the initial graphene growth and/or process. At $B$=-5T, $\rho_{xy}$ exhibits clearly quantized plateaus at $h/2e^2$ and $h/6e^2$ accompanied with the minimum in $\rho_{xx}$, and the higher order plateaus are developing as similarly reported by sweeping the external $B$ field and fixing charge density [10-13]. Here $h$ is the Planck constant and $e$ is the elementary electron charge. Such half-integer QHE, with quantization level at $h/4(n + 1/2)e^2$, with $n$ the Landau index, is the hallmark transport feature for monolayer graphene [3,23]. Observation of half-integer QHE is one of the important demonstrations that the epitaxial graphene on SiC and the exfoliated single-layer graphene are governed by the same relativistic physics with Dirac fermions as transport carriers, and that monolayer graphene can be formed on SiC substrate [10-13].

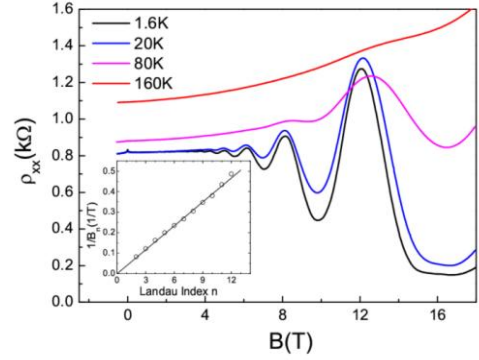

Fig. 3 Temperature dependence of $\rho_{xx}$ for device 1117A at $V_g$=1.8V from 1.6K to 160K. Inset: Landau plot of the maximum of the SdH oscillations up to 18T at 1.6K. $n$ is the Landau index and $B_n$ is the magnetic field at the corresponding maxima of the oscillations. The circles are the experimental data and the solid line is the linear fitting. The zero y intercept indicates the anomalous Berry's phase of π.

Since applying negative bias can dramatically decrease the carrier density and make the low filling factor ($v$=4$n$+2) QHE more easily visible, we keep the device biased at -5V, which corresponds to an electron density of 2.2×$10^{11}$ cm$^{-2}$ and mobility of 3250 cm$^2$/Vs at 420mK, and measure the temperature dependence of $\rho_{xx}$ and $\rho_{xy}$ as shown in Fig. 2. We note that $\rho_{xx}$ does not fully vanish at high magnetic fields even at the lowest temperature of 420mK due to the inhomogeneity of the epitaxial graphene within the 10×22 μm$^2$ device area and the defect-induced scattering that broadens the Landau level. Nevertheless, the $n$=0 quantum-Hall plateau and the corresponding diagonal resistivity minimum are still very pronounced within *magnetic field intervals of 15 Tesla (from 3T to the highest measured field 18T)*, even at *temperatures as high as 70K*. For 2.2×$10^{11}$ cm$^{-2}$ 2DES at $B$=18 T, the filling factor $v$ reaches as low as 0.5, deeply into the insulating phase if it exhibits. Possibly relevant to our observations, recent work by T.J.B.M. Janssen et al. points out that anomalously strong pinning of the $v$=2 could exist in epitaxial graphene due to the charge transfer between surface-donor states in SiC and graphene [24].

With further engineering of the device fabrication as well as graphene synthesis, it may become possible to observe and maintain the *n*=0 quantum Hall plateau on epitaxial graphene at room temperature [25]. Room temperature QHE on epitaxial graphene could be an interesting topic as a quantum resistance standard for metrology applications [11,24].

B.  Landau plot and Landau level separation

To further understand the underlying physics, the Landau plot, i.e. the Landau index vs. the inverse of the magnetic field, was investigated as shown in the inset of Fig. 3. Here we start with sample 1117A which has a much higher electron density of $2.4 \times 10^{12}$ cm$^{-2}$, determined by SdH oscillations and the Hall slope. It is more suitable for the Landau plot since more pronounced SdH oscillations are resolved, as shown in Fig. 3. The Landau index data was taken at the maximum of the SdH oscillations up to 18T at 1.6K. The open circles are the experimental data and the solid line is the linear fitting. The zero intercept at the x-axis indicates the anomalous Berry's phase of π. It is interesting to note that a remnant of the SdH oscillations can still be observed at 160 K at this quality of epitaxial graphene samples.

The damping of the SdH oscillation amplitudes in Fig. 3 is caused by thermal broadening of Landau level density of states. The temperature dependence of the relative peak amplitudes in graphene is given by [26]:

$$A_n(T)/A_n(0) = t_k / \sinh t_k \quad (1)$$

where $A_n(T)$ is the peak amplitude of the *n*th SdH peak at temperature *T*, and $t_k = 2\pi^2 k_B T / \Delta E(B)$ with:

$$\Delta E(B) = E_{n+1}(B) - E_n(B) \quad (2)$$
$$\Delta E(B) = (\sqrt{n+1} - \sqrt{n}) v_F \sqrt{2eB\hbar} \quad (3)$$

the Landau level separation. Here, $k_B$ is the Boltzmann constant and $v_F$ is the Fermi velocity of the carriers in graphene.

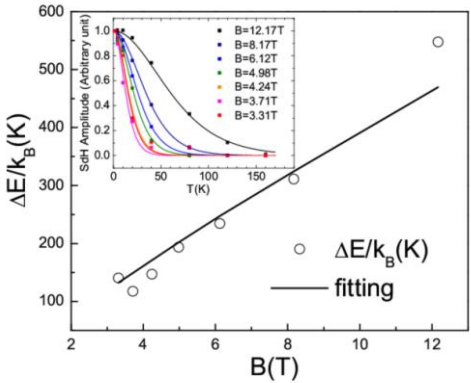

Fig. 4 (a) Fitting to the Fermi velocity $v_F$ according to the Landau level separation. (b) Relative SdH peak fitting according to Eq. 1 for different temperatures at each peak.

The experimental values of the Landau level separation were determined by fitting the normalized SdH peak amplitude according to Eq. 1 as a function of temperature, as shown in the inset of Fig. 4. Plotting the normalized amplitude allows easier graphical comparison between the data taken at different magnetic fields. The corresponding $\Delta E(B)$ versus *B* is plotted as the open circles in Fig. 4. The dotted line is the fitting of $\Delta E(B)$ versus *B* according to Eq. 3. The electron Fermi velocity $v_F$ in graphene is found to be $1.01 \times 10^8$ cm/s from the fit. Similar analyses were also carried out for devices 1118A and 1118B, with $v_F = 1.05 \times 10^8$ cm/s and $9.94 \times 10^7$ cm/s. These results are around the accepted value of $v_F = 1.00 \times 10^8$ cm/s for graphene.

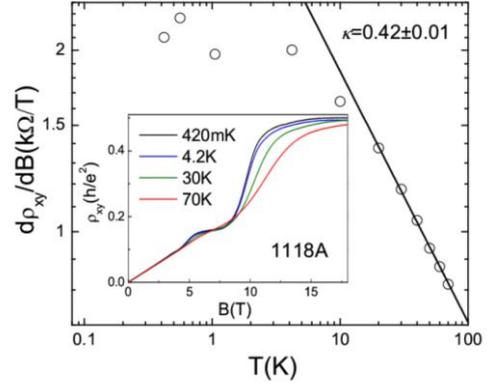

Fig. 5 The derivative ($d\rho_{xy}/dB$) at the transition's critical point *B*=8.4T as a function of temperature for the n=1 plateau to n=0 plateau transition of sample 1118A. The open circles are the experimental data and the solid line is the fitting for the highest 6 temperature points. Inset: $\rho_{xy}$ versus *B* of sample 1118A from 420mK to 70K at $V_g$=0V.

C.  Universal scaling in plateau (*n*=0) to plateau (*n*=1) transition

In scaling descriptions of quantum-Hall inter-plateau transitions, the localization length ξ, i.e. the spatial extension of the electron wave function, diverges as a power law $\xi \sim (E-E_c)^{-\gamma}$ with a universal critical exponent γ of 2.4 [27,28] where $E_c$ is the energy of a Landau level center. Since $E_c$ varies with *B*, the derivative $(d\rho_{xy}/dB)^{max}$ at the critical point of the plateau-to-plateau transition and the $\rho_{xx}$ peak width, which is defined as the distance $\Delta B$ between the two extremes in $d\rho_{xx}/dB$, provide the experimental measure of the delocalization phenomenon in the integer quantum Hall regime [27,28]. It has been found that, below a certain characteristic temperature $T_{sc}$, $1/(d\rho_{xy}/dB)^{max}$ and $\Delta B \sim T^\kappa$ where $\kappa \approx 0.42$, is universal for conventional 2D systems with short-range alloy disorders [27-29].

In exfoliated graphene [30], $\kappa \approx 0.37$ has been reported for the half width in $\rho_{xx}$ of the first Landau level and $\kappa \approx 0.41$ for the derivatives $(d\rho_{xy}/dB)$ at the critical point of the plateau (*n*=0) to plateau (*n*=1) transition. In a plateau-insulator quantum phase transition passing the last plateau in exfoliated graphene, $\kappa\approx0.58$ is obtained [31]. Here, we demonstrate that the ($n$=0)-to-($n$=1) inter-plateau transition in epitaxial graphene also exhibits behavior consistent with scaling in the same universality class as (1) the same transition [30] in the mechanically exfoliated graphene, and (2) various inter-plateau transitions in semiconductor-hosted 2D electron systems [27-29]. Unfortunately, we are not able to investigate the scaling behavior for higher Landau level transitions, since they are not sufficiently well-developed in our samples.

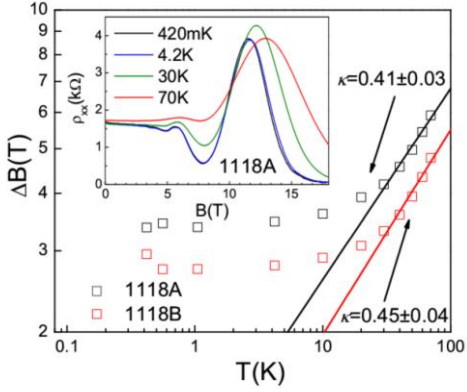

Fig. 6 $\rho_{xx}$ peak width $\Delta B$, the distance between the two extremes in the derivative ($d\rho_{xx}/dB$) of the first Landau level as a function of temperature for sample 1118A and 1118B. The hollow symbols are the experimental data and the solid lines are the fitting of the highest 5 temperature points. Inset: $\rho_{xx}$ versus $B$ of sample 1118A from 420mK to 70K at $V_g$=0V.

The inset of Fig. 5 shows the $\rho_{xy}$ vs $B$ from sample 1118A at several temperatures from 420mK to 70K, with $V_g$=0. Since this specific sample shows the best plateau-to-plateau transition, it is a good starting point to analyze the scaling behavior. Fig. 5 shows the temperature dependent derivative $(d\rho_{xy}/dB)^{max}$ at the critical point $B$=8.4 T of the ($n$=0)-to-($n$=1) inter-plateau transition of the same sample. The open circles are the experimental data and the solid line is the fit for the highest six temperatures. The slope shows $\kappa$= 0.42±0.01. Here (and in the $\kappa$ values stated below) the error is determined from scatter about the linear fit, and neglects uncertainty due to the choice of the temperature range of the fit. Though the range of temperature covered by the fit is limited, the agreement with the accepted universal value of 0.42 is remarkable.

We also study the temperature dependence of the peak width, $\Delta B$, in $\rho_{xx}$, with the inset of Fig. 6 showing the $\rho_{xx}$ vs $B$ of sample 1118A at temperatures from 420mK to 70K. Fig. 6 shows the extracted peak width $\Delta B$ as a function of temperature for both sample 1118A and 1118B. The open symbols are the experimental data and the solid lines are the fits for the highest five temperatures. The slope shows $\kappa$= 0.41±0.03 for 1118A and $\kappa$= 0.45±0.04 for 1118B again in excellent agreement with the standard value for $\kappa$.

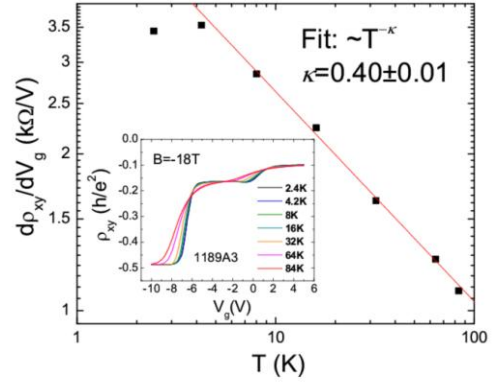

Fig. 7 The derivative ($d\rho_{xy}/dV_g$) at the transition's critical point $V_g$=-5.95V as a function of temperature for the $n$=1 plateau to $n$=0 plateau transition of sample 1189A3. The filled squares are the experimental data and the solid line is the fitting for the highest 6 temperature points. Inset: $\rho_{xy}$ versus $V_g$ of sample 1189A3 from 2.4K to 84K at $B$=18T.

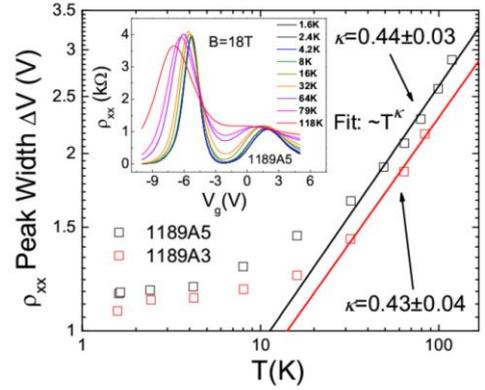

Fig. 8 $\rho_{xx}$ peak width $\Delta V$, the distance between the two extremes in the derivative ($d\rho_{xx}/dV_g$), of the first Landau level as a function of temperature for sample 1189A3 and sample 1189A5. The solid squares are the experimental data and the solid lines are the fitting of the highest 5 temperature points. Inset: $\rho_{xx}$ versus $V_g$ of sample 1189A5 from 1.6 K to 118 K at $B$=18T.

We also investigate the temperature dependence of plateau-to-plateau transitions with gate sweeps for two samples, 1189A3 and 1189A5. Diagonal resistance, $\rho_{xx}$, and Hall resistance, $\rho_{xy}$, as a function of top-gate voltage at $B$=18T for different temperatures are plotted in the insets of Figure 7 and Figure 8, respectively. The derivative at the transition critical point, $(d\rho_{xy}/dV_g)^{max}$, and the extracted $\rho_{xx}$ peak width, defined as the distance $\Delta V$ between the two extremes in $d\rho_{xx}/dV_g$, are plotted vs temperature in Figure 7 and Figure 8 for the samples at a fixed 18T magnetic field. Again considering the highest temperature points, Figure 7 con-firms a power-law behavior following $(d\rho_{xy}/dV_g)^{max} \sim T^{-\kappa}$ with measured $\kappa$=0.40±0.04 at $V_g$=-5.95V, while Figure 8 has fits to $\Delta V \sim T^\kappa$ obtaining $\kappa$=0.44±0.04 for sample 1189A5 and $\kappa$=0.43±0.04 for sample 1189A3 as shown in. These results confirm the universality of a critical quantum Hall scaling in epitaxial graphene with gate sweeps as well.

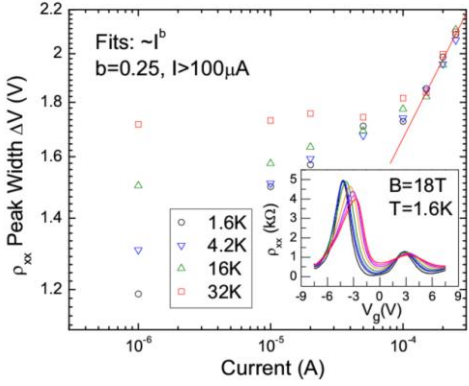

Fig. 9 $\rho_{xx}$ peak width $\Delta V$, the distance between the two extremes in the derivative $(d\rho_{xx}/dV_g)$, of the first Landau level as a function of current for sample 1189A5 measured at four different temperatures. The empty symbols are the experimental data and the solid lines are the fitting of the three groups of data points at $I$>100μA. Inset: $\rho_{xx}$ versus $V_g$ of sample 1189A5 with different $I$ from 250 μA down to 1μA at T=1.6K and B=18T.

In all the plots of vs temperature in Figures 5-8, the transition width, as measured by maximal $\rho_{xy}$ derivatives or $\rho_{xx}$ peak widths, has scaling power-law behavior at high temperature, with saturation of the width at low temperature, less than ~10 K. Similar low temperature flattening of the transition width vs temperature was also reported in ref. 30. The interpretation of this sort of saturation [28,29] is that the localization length ξ is approaching another length scale intrinsic to the sample. In refs 28 and 29 this cut-off length was shown to be the sample size, by comparison of samples of several sizes. In the graphene case sample size is a possible cutoff length as well, but it is also possible that the gate dielectric thickness may be the cut-off length, since the Coulomb interaction crosses over to a dipolar interaction on length scales larger than the dielectric thickness. Scaling behavior can be affected by the form of the interaction [32]. Also, as in ref 30, the interpretation of our data as in agreement with the accepted value of κ requires the fourfold degeneracy of the graphene level to have no effect on the universality class as in some cases reported in ref [33], but unlike the case of spin-degenerate inter-plateau transitions in ref 34.

Finally, Fig. 9 shows the effect of current on the inter-plateau transition in epitaxial graphene. DC measurements were used when studying the current scaling. The current dependent measurement is correlated with the electrical field heating effect, where the input current ($I$) is proportional to the electrical field ($E$) in the sample. When an electrical field $E$ is applied, energy is being put into the sample. The effective temperature ($T_e$) of the electron gas is then different from the helium bath temperature ($T_b$). Beyond a critical $E$, the measured $\rho_{xx}$ and $\rho_{xy}$ reflect the temperature $T_e$ instead of $T_b$. This kind of current scaling behavior has been observed in conventional 2D systems with $(d\rho_{xy}/dB) \sim I^{-b}$ where $b$=0.23±0.02 and $b=\kappa/2$, implying $T_e \sim I^{0.5}$ [35]. The inset of Figure 9 shows the $I$ dependence of $\rho_{xx}$ at $T_b$=1.6K and $B$=18T. Comparing the insets of Figure 8 and Figure 9, the $T$ and $I$ dependent evolutions of $\rho_{xx}$ are qualitatively the same. In Figure 9, we plot the $I$-dependent $\rho_{xx}$ peak width $\Delta V$ measurements at $T_b$=1.6K (circles), 4.2K (down triangles), 16K (up triangles) and 32K (squares). Current scaling similar to the semiconductor-hosted 2D system, is also observed in epitaxial graphene, with b≈0.25 from fitting the three groups of data at $I$ >100 μA. However, less than the critical current of 100μA, the current scaling changes as a function of the bath temperature $T_b$. As $T_b$ is increased, the $\rho_{xx}$ peak width becomes less sensitive to the current $I$, becoming largely insensitive at $T_b$=32K. The results in Figure 9 are consistent with the universal scaling and $T_e \sim I^{0.5}$ in the epitaxial graphene for sufficiently large $I$.

IV. CONCLUSION

A high-k gate stack on epitaxial graphene is realized by inserting a fully oxidized nanometer thin aluminum film as a seeding layer followed by an atomic-layer deposition process. The resulting device has convenient density tenability, and adequate mobility to exhibit strong quantum Hall effects even at high temperatures. The experiments confirm that the universal scaling in quantum-Hall plateau to plateau transition also holds for epitaxial graphene on SiC (0001) as previously observed in conventional 2D systems and exfoliated graphene.

ACKNOWLEDGMENTS
The authors would like to thank G. Jones, J. Park, T. Murphy, and E. Palm at National High Magnetic Field Laboratory (NHMFL) for experimental assistance. Part of the work on graphene is supported by NRI (Nanoelectronics Research Initiative) through MIND (Midwest Institute of Nanoelectronics Discovery), DARPA, and Intel Corporation. NHMFL is supported by NSF Grant Nos. DMR-0084173 and ECS-0348289, the State of Florida, and DOE.